\documentclass[pdflatex,sn-mathphys]{sn-jnl}

\usepackage{graphicx}
\graphicspath{ {./images/} }

\jyear{2021}%

\theoremstyle{thmstyleone}%
\theoremstyle{thmstyletwo}%

\theoremstyle{thmstylethree}%

\begin{document}

\title[Article Title]{Integrated Photonic Platforms for Quantum Technology: A Review}


\author*[1]{\fnm{Rohit} \sur{K Ramakrishnan}}\email{rohithkr@iisc.ac.in}

\author[1]{\fnm{Aravinth} \sur{Balaji Ravichandran}}\email{aravinthb@iisc.ac.in}

\author[1]{\fnm{Arpita} \sur{Mishra}}\email{arpitamishra@iisc.ac.in}

\author[2]{\fnm{Archana} \sur{Kaushalram}}\email{archana@iisc.ac.in}

\author[3]{\fnm{Gopalkrishna} \sur{Hegde}}\email{gopalkrishna@iisc.ac.in}

\author[1]{\fnm{Srinivas} \sur{Talabattula}}\email{tsrinu@iisc.ac.in}

\author[4]{\fnm{Peter} \sur{P Rohde}}\email{peter.rohde@uts.edu.au}

\affil*[1]{\orgdiv{Department of Electrical Communication Engineering}, \orgname{Indian Institute of Science}, \orgaddress{\city{Bengaluru}, \postcode{560012}, \state{KA}, \country{India}}}

\affil[2]{\orgdiv{Department of Instrumentation and Applied Physics}, \orgname{Indian Institute of Science}, \orgaddress{\city{Bengaluru}, \postcode{560012}, \state{KA}, \country{India}}}

\affil[3]{\orgdiv{Centre for BioSystems Science and Engineering}, \orgname{Indian Institute of Science}, \orgaddress{\city{Bengaluru}, \postcode{560012}, \state{KA}, \country{India}}}

\affil[4]{\orgdiv{Centre for Quantum Software and Information}, \orgname{University of Technology Sydney}, \orgaddress{\city{15 Broadway Ultimo}, \postcode{2007}, \state{NSW}, \country{Australia}}}


\abstract{Quantum information processing has conceptually changed the way we process and transmit information. Quantum physics, which explains the strange behaviour of matter at the microscopic dimensions, has matured into a quantum technology that can harness this strange behaviour for technological applications with far-reaching consequences, which uses quantum bits (qubits) for information processing. Experiments suggest that photons are the most successful candidates for realising qubits, which indicates that integrated photonic platforms will play a crucial role in realising quantum technology. This paper surveys the various photonic platforms based on different materials for quantum information processing. The future of this technology depends on the successful materials that can be used to universally realise quantum devices, similar to silicon, which shaped the industry towards the end of the last century. Though a prediction is implausible at this point, we provide an overview of the current status of research on the platforms based on various materials.}




\maketitle

\tableofcontents

\section{Introduction}\label{sec1}

\textit{Quantum} is the buzzword of the decade. The advent of quantum technologies that range from quantum computers and communications to quantum satellites and sensors is considered the most prominent technological advance of the decade by industry experts and academicians. These devices claim unbreakable security and possess enormous computing power that is unachievable by their classical counterparts. Quantum technologies have passed many milestones in the previous decade, ranging from quantum supremacy to quantum satellites enabling ground-to-earth quantum communication. However, the major limitation of these experiments is that they are conducted primarily in a controlled laboratory environment and quantum optic benches. To be a viable technological alternative and available for the public with the advantages mentioned above, we should be able to manufacture quantum devices on a massive scale without losing their properties. There is still a long way to the target from a practical perspective. \\

We are still very much in the era of nanotechnology, courtesy of the materials and fabrication techniques. To have quantum devices replacing or complementing the current technology, we need much higher levels of scalability, which remains the primary hurdle for their extensive application. Though we have started to witness quantum devices at the chip level, the consensus from the experiments is that photons are the better carriers for quantum information, not electrons. In this context, photonic integrated circuits (PIC) provide a possible solution with a high density of integrated components and enhanced performance. PIC offers a wide range of functionality that includes generating and detecting non-classical states for quantum information to amplification, modulation, frequency translation, switching, and routing quantum signals. According to industry experts, it can be benefited from the developments in classical integration technologies in photonics.  \\

The significant challenges in developing integrated photonic platforms for quantum applications lie in different quantum modalities' requirements. We will need photons to serve as qubits and interface the qubits. Quantum technological applications rely heavily on the efficient generation and detection of entangled photons. Further requirements include minimum propagation loss, optical coherence control and the ability to operate at various wavelengths. Though integrated photonic technology successfully demonstrates many quantum functions, the requirements of photonics n massive scale quantum systems are still not completely defined. The next major step will be integrating with electronics to provide power, control, and I/O with the classical world. As the qubits' density increases, the required interconnects and integration levels also rise. A critical factor will be to make it scalable and, at the same time, shield the quantum components from the electromagnetic interference  (EMI) produced by the electronic components. \\

The era of micro and nanotechnology has been made possible by the rigorous developments in silicon-based process technology for mass production. Silicon process technology developed during the last fifty years has become an industry standard with a well-matured nanoscale fabrication, packaging technology and mass production capabilities. Hence silicon-based technology has become the most sought-after choice for fabricating platforms for quantum information processing. One of the main advantages is that they can piggyback off existing CMOS fabrication technology, allowing the infrastructure of an existing multi-trillion dollar economy to be utilised. However, there have been many limitations to silicon-based platforms. One of the main issues is the photon losses through the device, which has inspired researchers worldwide to look for alternate materials with specific advantages over silicon. The discovery of new materials with interesting electronic and optical properties has given rise to new platforms which can also support the implementation of quantum functionalities. The developments in thin-film technologies and quantum dots have accelerated these new possibilities. Given the hybrid nature of many integrated platforms, it requires further research to decide on the materials that can successfully be used to design and implement quantum devices across platforms. Various materials have been explored to develop efficient quantum device platforms in recent years. We are still a long way from achieving a standard platform that will dominate the quantum computers like silicon which dominated the classical computers. This paper surveys the latest advancements in various integrated photonic platforms based on diverse materials being explored for quantum technology \cite{Zhong2020,Wang2020,Moody2022,Pelucchi2022}.

\section{Silicon based Platform}\label{sec2}

Silicon is the material that has revolutionised the electronics industry in the past century with its electrical and thermal properties and availability. The advantages of a silicon based platform are well-established CMOS processes and foundries, high component density, ability to operate at telecom wavelength and the nonlinear properties which make it attractive for many integrated photonic applications. Therefore silicon has been the first choice for many demonstrations of building blocks for photonic quantum computing. More than any other, silicon based platforms have seen expeditious growth in the number of photons and integrated components in the last decade. \\

One of the earliest testbeds for quantum devices was silica-based platforms. High fidelity on-chip quantum logic operations were demonstrated in glass-based waveguides. Though it includes the generation and manipulation of entangled photons, multi-particle quantum walks and boson sampling, silica-based platforms are limited by the scalability of complex circuits and their functionality. \cite{7479523,Politi2008,Peruzzo2010,Shadbolt2012,Spring2013,Tillmann2013} 

\subsection{Photon sources}\label{subsec1}

The first task for quantum information processing is state preparation. This involves generating single photons and photon pairs as photons are the most widely accepted quantum information carriers. The error rates and scalability determine the quality of such sources. Ideally, photons generated from such a source should be pure, deterministic and indistinguishable, which implies that they must be identical in every degree of freedom except time. The efficiency of a photon source is denoted by \textit{p}, the probability that a photon is emitted. \\

Spontaneous four wave mixing (SFWM) is the most commonly used photon generation method in silicon based platforms. In this method two photons from a bright pump field are converted to energy and momentum conserving photon pair by $\chi^{(3)}$ non-linearity in crystalline silicon. They are called signal and idler photons and the quantum state generated is called a squeezed vacuum state.
\begin{equation}
 \vert\psi\rangle = \sqrt{1-\mid \xi \mid^2}\sum_{n=0}^{\infty}(-\xi)^n\vert n,n \rangle_{si}
\end{equation}

where $\xi =ie^{iarg(\zeta)} \tanh {\mid \zeta \mid}$ is the squeezing parameter and $\zeta \propto I^2$ for pump power $I$. For small $\zeta$, $\vert\psi\rangle$ is dominated by vacuum, with a two photon component. Four-photon and above components can be made small by controlling the pump laser power. In non linear sources, generally there is always a trade off between photon number impurity (multi photon noise) and number of photons (brightness). In such sources brightness ranges from $0.05 < p < 0.1$ depending on their applications.\\ 

The purity $P \in [0, 1]$ of a quantum state is a measure that indicates how much a photon is entangled with any other. It is calculated as $tr (\rho^2)$ for a state with density matrix $\rho$. Pure states $(P = 1)$ are totally not entangled, while mixed states $P < 1$ entangled. Purity is also used to describe the number-basis deviation of a photonic state while considering the photon sources. For single photon sources, entanglement is not a significant issue, but for nonlinear photon-pair sources, purity refers to a lack of entanglement \cite{Adcock2021}. The Joint Spectral Density (JSA) of a photon pair source decides the purity of its spectrum. This is obtained either by chromatic group dispersion time-of-flight measurements or stimulated four-wave mixing. Recently researchers have successfully engineered interference between the output of stimulated emission tomography with a coherent phase reference beam, which enables them to measure phase and thereby get complete information on JSA. Massimo Borghi has successfully measured the JSA of a silicon ring resonator by using a straight waveguide
as a nonlinear phase reference \cite{Eckstein2014,Silverstone2015,Avenhaus2009,Erskine2018,Borghi2020}. \\

To make large-scale photonic circuits in silicon a reality, we need repeatable sources of SFWM. Jay Sharping et al. demonstrated that a parameter required for a waveguide to generate correlated photon pairs is its length \cite{Sharping2006}. However, the waveguide sources are limited by spectral purity.\\

Ring resonators are another approach for generating photons in silicon photonics. We can engineer the resonances in them to produce more symmetric JSA and enhance SFWM processes for specific wavelengths with a spectral purity up to 0.93 \cite{Clemmen2009,Helt2010,Vernon2017,Silverstone2015}. The scalability of ring resonator sources is restricted by cross talk and environmental factors like temperature variations. Llewellyn et al. recently demonstrated high-quality ring resonator sources on-chip across a large set of phase configurations \cite{Llewellyn2020}. Carolan et al. has shown that resonances can be stabilised to a value less than that of 1\% of linewidth drift, using a photodiode and a PID feedback loop, which has proven to be a scalable approach \cite{Carolan2019}. \\

For higher scalability levels that require purity levels greater than 0.93, either multi-ring structures or asymmetric MZI coupling is used to shape the signal and idler resonances. Experimentally, this has been realised in silicon platforms by generating entanglement between photonic qutrits with purities up to 0.95 and heralding efficiency of 52\% \cite{Liu2020,Lu2020}. Manipulation of the photon dynamics inside the ring can also be used to improve spectral purity for high-level scalability. If both halves of the pulse are opposite in phase, it de-excites the pump resonance, thus reducing the pump resonance quality factor \cite{Christensen2018,Vernon2017}.With an MZI coupled resonator, using stimulated emission tomography, a spectral purity of 0.98 was obtained by Burridge et al. in 2020 \cite{Burridge2020}. Photon pairs are also produced using multi-dimensional arrays of silicon ring resonators that exhibit topological effects, though it is limited by the size and presence of other resonant modes \cite{Mittal2018}. Paesani et al. demonstrated a multimode waveguide source with 0.99 spectral purity and 0.99 spectral overlaps. They split a pump pulse into two waveguides, delayed one by 13 ps, and used a multimode coupler to transform the waveguides into a wider cross-section waveguide. They claimed methods to improve the parameters and make the fabrication with 4 nm lithographic precision \cite{Paesani2020}.

\subsection{Passive Components}\label{subsec2}

The promise of silicon technology is the integration of whole quantum photonic circuits in a single chip. This includes passive components for quantum information processing like fiber-to-chip couplers, 2×2 couplers, cross intersections, multimode and polarisation components, filters, delay lines, etc.\cite{Adcock2021} \\

Two types of fiber-to-chip couplers are mainly used in silicon based platforms- lateral couplers and vertical couplers. Laterals couplers are based on spot size converters (SSCs) which have low polarisation dependency and higher coupling bandwidth. The fabrication of SSCs with lower noise is highly challenging, as the lowest loss reported is less than 1 dB for coupling with lensed fibers and single mode fibers \cite{Pu2010,Chen2010,5432953}. Grating couplers are generally polarisation dependent, and bandwidth limited, limiting their capacity to handle quantum information. Two-dimensional grating couplers that can couple orthogonally polarized photons into separate waveguides are used to overcome this but at the cost of signal loss. Attempts have been made to overcome this situation by optimizing the waveguide height and partial etch depth and apodization of fill factor and grating period \cite{Llewellyn2020,Taillaert2004,Van4383198,Wang2016}. Notaros et al. demonstrated a grating coupler with 0.36 dB loss using 45 nm process with two partial etch depths and using a Bloch-Floquet band-structure type optimisation \cite{Notaros2015,Notaros2016}. Luo et al. has shown a 3D polymer coupler with a coupling loss as low as 1 dB \cite{Wang2014}.\\

Mach-Zehnder interferometers (MZIs) are prominent devices that manipulate photons and thereby realise quantum gates. In silicon platforms, they consist of two modes joined by multimode interferometers or directional couplers with an enclosed phase modulator, forming a programmable interferometer. Using MZIs with mismatched internal path lengths, sinusoidal wavelength filters are realised to split single photon wavelengths. MZIs are advantageous in terms of fabrication tolerance and low insertion losses \cite{Dumais2016}. Though directional couplers have standard propagation losses, they are less compared to MZIs, which the subwavelength structure can reduce \cite{Halir2012,Xie2019}.\\

Qubits are realised in silicon waveguides by either polarisation modes or spatial modes. Polarisation dependent directional couplers that exploit the differing coupling constants of TE and TM modes can be used to multiplex the polarisation modes. This can achieve bandwidths of more than 100 nm \cite{Fukuda2006}. Directional couplers can also be used to realise mode de-multiplexers, the state-of-the-art being eight modes \cite{Dai2012,Wang2014}. The sensitivity of cross-mode coupling to fabrication tolerance can be alleviated through a taper in the coupling region \cite{Ding2013}.\\

In path encoded qubit circuits, waveguides have to cross each other, which could cause cross talk, and thus qubit errors \cite{Wang2017}. New structures with insertion loss less than 0.2 dB and crosstalk less than -35 dB have been proposed based on MZI structures, shaped and tilted waveguides, sub-wavelength gratings, Bloch mode structures and by the inverse design method \cite{Wang2018,1715394,Fukazawa2004,Sanchis2009,Ma2013,Xu2011,Kim2014,Bock2010,Liu2014,Yu2019,Zhang2013}. A detailed review of silicon waveguide crossings has been published by Wu et al. \cite{Wu2020}\\

Silicon-based platforms typically have propagation losses of the order 2 dB/cm, which restricts scaling up for quantum technological applications. This has been reduced to 2.7 dB/m with the advancements in fabrication technology like oxidation, etchless fabrication, hydrogen thermal annealing and shallow waveguides. Hybrid integration with other platforms like Silicon Nitride and small-angle wedge silica waveguides have shown promise of further reduction of these losses up to 0.123 dB/m and 0.037 dB/M, respectively \cite{Adcock2021}.

\subsection{Modulators and Switches}\label{subsec3}

The only modulators available in the silicon photonic platforms with low-loss and small footprint are thermo-optic phase modulators. These are implemented either by metal heating elements suspended close to the waveguide or n-type or p-type silicon connected with the waveguide. The latter has higher bandwidth but a higher loss also. Also, the operating frequencies of thermo-optic modulators fall in the KHz range and this is a huge limitation as far as scalability is concerned. Plasma-based modulators and switches are also popular since their operating frequencies are in the GHz range and compatible with the existing CMOS technology. But they are limited by lower efficiencies, higher power consumption and large device footprint \cite{Adcock2021,Harris2014,Fang2011,Watts2013}\\

The best modulators for quantum information processing are primarily available on non-silicon based platforms like Lithium Niobate with a loss of around 2.7 dB/m. The challenge with the Lithium Niobate platform is its efficiency in coupling. This can be overcome by making hybrid devices with silicon using inverse tapers, which can do efficient coupling of light with nearly zero loss \cite{He2019}. This has also been demonstrated using silicon nitride \cite{Boynton2020}. Recent experimental advances have established that lithium Niobate on silicon has the potential to higher clock rate electronic-photonic integration at CMOS compatible voltages with losses as low as 0.5 dB \cite{Wang2018}.\\

Pure silicon, under strain, will break the symmetry of its lattice structure, which will induce a non-linearity $\chi^{(2)}$. Using this phenomenon, phase modulation up to 1 GHz has been demonstrated in silicon waveguides, at cryogenic temperatures also \cite{Jacobsen2006,Timurdogan2017,Chakraborty2020}. In these devices, strong electric fields are created by p-i-n junctions, which distort the crystal lattice and break its symmetry—through free carrier absorption while keeping the phase dependant loss as low as 1 dB/cm.

\subsection{Detectors}\label{subsec4}

A primary resource for quantum information processing is the detector. Silicon photonics is opaque for wavelengths below 1100 nm and sensitive to the telecommunication bands, i.e. 1530-1565 nm, as they are optimised for single-mode fiber transmission. Detection in the telecom band is mainly done by InGaAs and InP detectors. They are cheaper with efficiency lower than 10\%. The count rates are measured in the MHz range, and dark count rates are typically around 100 Hz. hence, these can be used only in short and medium link quantum communication scenarios as their losses are affordable \cite{Zhang2015}. \\

Superconducting nanowire single photon detectors (SNSPDs) are excellent candidates for photon detection due to their efficiency ranging up to 95.5\% and dark count rates less than 100 Hz. Commercially available SNSPDs typically work in the 4K temperature range with recovery times that enable more than 107 detections per second. Though these detectors have taken quantum photonics in silicon forward, their inability to work in higher temperatures negatively impacts cost. Also, these detectors are off-chip and fiber-coupled, which results in low efficiency. On-chip integration of SNPSDs is still a challenging topic \cite{Natarajan2012, Rosenberg2013, Zhang2017}. SNSPDs in travelling wave configuration with almost 100\% efficiency have been demonstrated in silicon waveguides which holds great promise for integration \cite{Schuck2013, Akhlaghi2015}. Recent developments include detectors written using custom thin-film techniques on silicon photonic circuits. Though the grating couplet efficiency achieved is only -6 dB, this paves the way for higher scalability of SNSPDs in silicon waveguides \cite{Zheng2021}. It has also been demonstrated that single photon operations can be performed using electrically driven light sources by integrating silicon waveguide SNSPDs with on-chip light sources like LEDs and single photon emitting carbon nanotubes \cite{Buckley2017, Khasminskaya2016}. Commercial systems typically require detector recovery times of 50 ns, which can be achieved by reducing the long detector length required for efficient photon absorption. A method to achieve this is using silicon photonic crystal cavities. These cavities use short detectors with recovery times around 500 ps and also provide ultra-narrow wavelength specificity, which is helpful in frequency multiplexing \cite{Vetter2016, Julian2018}. Another limitation in quantum communication applications for the detector is the jitters or temporal resolution that form a bottleneck in ultra-fast temporal multiplexing. SPNSDs in silicon substrate have successfully overcome this limitation predominant in TFLN and BTO modulators. Korzh et al. demonstrated 3 ps resolution using 5 $\mu$m SNSPDs made from RF-bias sputtered niobium nitride films and cryogenic electronics. The performance was 2-3.5 times better than other materials like Tungsten silicide, and this low jitter was because of their quasi-amorphous small crystal structure \cite{Korzh2020}.\\

A prominent type of detector required in applications like Gaussian boson sampling and universal quantum gates are photon number resolving detectors. SNSPDs can achieve number resolution in monolithic devices by spatial, temporal and spectral multiplexing, though it has not been successful in a silicon platform \cite{Barry2002, Divochiy2008, Allman2015, Sahin2013, Fitch2013, Mattioli2016}. Since this technique is based on the processing of the electronic detector pulses, theoretically, they can be transferred to any platform. Cheng et al. demonstrated a broadband chip-scale single photon spectrometer in visible and infrared bands \cite{Cheng2019}. They produced a single photon camera with an area of 590 effective pixels in 286×193 $\mu$m$^2$. When it comes to higher scalability, SNSPDs readout has to be multiplexed because of the thermal budgeting constraints \cite{Gaggero2019}. Di Zhu et al. had demonstrated a single channel readout and photon number resolution across 16 detectors of all 136 single and two photon events \cite{Zhu2018}. Signal processing and multiplexing techniques in cryogenic temperatures have become the hot topic of research in the context of SNSPDs. An SNSPD array of 2 × 32 pixels has been demonstrated with row-column multiplexing.  \cite{Wollman2019}. The fiber out of the cryostat was used for optical data transfer by Marc de Cea et al., which shows capability for high-speed signal multiplexing \cite{MarcdeCea2020}.\\

Silicon-germanium single photon avalanche detectors are an alternative to the expensive and bulky cryogenic detectors since they can be waveguide coupled and have shown efficiency up to 38\% at 1310 nm at 125 K temperature. The drawback of these types of detectors are dark counts in the 104 range and timing resolution, which can be in the order of hundreds of picoseconds \cite{Martinez2017, Vines2019}. If further research can improve these parameters, we can have non-cryogenic devices for many quantum applications soon.\\

A challenge faced in most quantum photonic devices is removing pump light used to generate single photons before detection. Though off-chip elements mainly do this, on-chip removal of the pump will further improve single photon detectors in performance and integration. Most applications require over 100 dB of extinction—with minimal losses in the single photon bands. Quantum communications networks require filters to separate classical and quantum channels. Coupled resonator optical waveguides (CROWs) formed by cascaded microring resonators and cascaded asymmetric MZIs are currently used to achieve more than 50 dB extinction on a single chip. Pump light due to the grating couplers with efficiency as low as -5 dB was scattered into the chip substrate limiting the filtering ratio \cite{6527974, Piekarek2017}. \\

Gentry et al. have demonstrated single photon generation and pump rejection together on a single chip. They used four second-order ring resonator filters to isolate the signal and idler bands from the pump \cite{Gentry2018}. An extinction ratio of more than 70 dB has been achieved in silicon nitride using cascaded grating-assisted contra-directional couplers. In comparison, an extinction ratio of more than 80 dB has been achieved with third-order ring resonators \cite{Nie2019, Huffman2017}. Also, an extinction of more than 100 dB was achieved by using two chips \cite{6527974, Piekarek2017, Kumar2020}.\\

\subsection{Beyond Silicon}\label{subsec5}

Even though silicon provides a straight route to large-scale manufacturability, such a platform has the drawbacks of being homogeneous and entirely dependent on a single material for realising quantum technology's core functions. The photon generation, as we discussed, is spontaneous, and there is no direct photon-photon interaction. Though probabilistic measurements and feed-forward operations compensate for the lack of deterministic methods to generate, store, and entangle photons, these methods are yet to be perfected. A heterogeneous platform like SiN, LN, or AlN with broad transparency would be more advantageous for integrating the hardware elements ranging from photon sources to quantum memories. However, the large scale fabrication of such devices is also a challenging area \cite{Moody2022}. \\

If silicon-based platforms have to be the primary choice for quantum technological applications, they need to overcome challenges like operating only in cryogenic temperatures for detection and addressing the issues in interfacing electronics and photonics. Tasker et al. recently demonstrated the detection of squeezed light using a silicon photonic integrated chip wire-bonded directly to a silicon electronics die. The total device capacitance was curtailed by packing the components closer. This is the highest bandwidth of shot-noise limited detector performance reported \cite{Tasker2021}. The significant limitations in monolithic electronic–photonic integrated circuits are limited availability of high-performance components and lack of simultaneous expertise.\\

Carrier injection modulation and thermal phase shifting are the primary switching mechanisms used in silicon. These techniques suffer from phase-dependent losses and limited speed, respectively. Since silicon does not exhibit second-order nonlinearity, either an external field has to be applied or materials capable of supporting fast electro-optic (EO) switchings, such as LN or barium titanate, has to be integrated by engineering high-efficiency coupling to silicon for fast switching \cite{Tasker2021, He2019, Eltes2020}.\\

\begin{figure}[ht]
	\centering
	\includegraphics[width=0.9\linewidth]{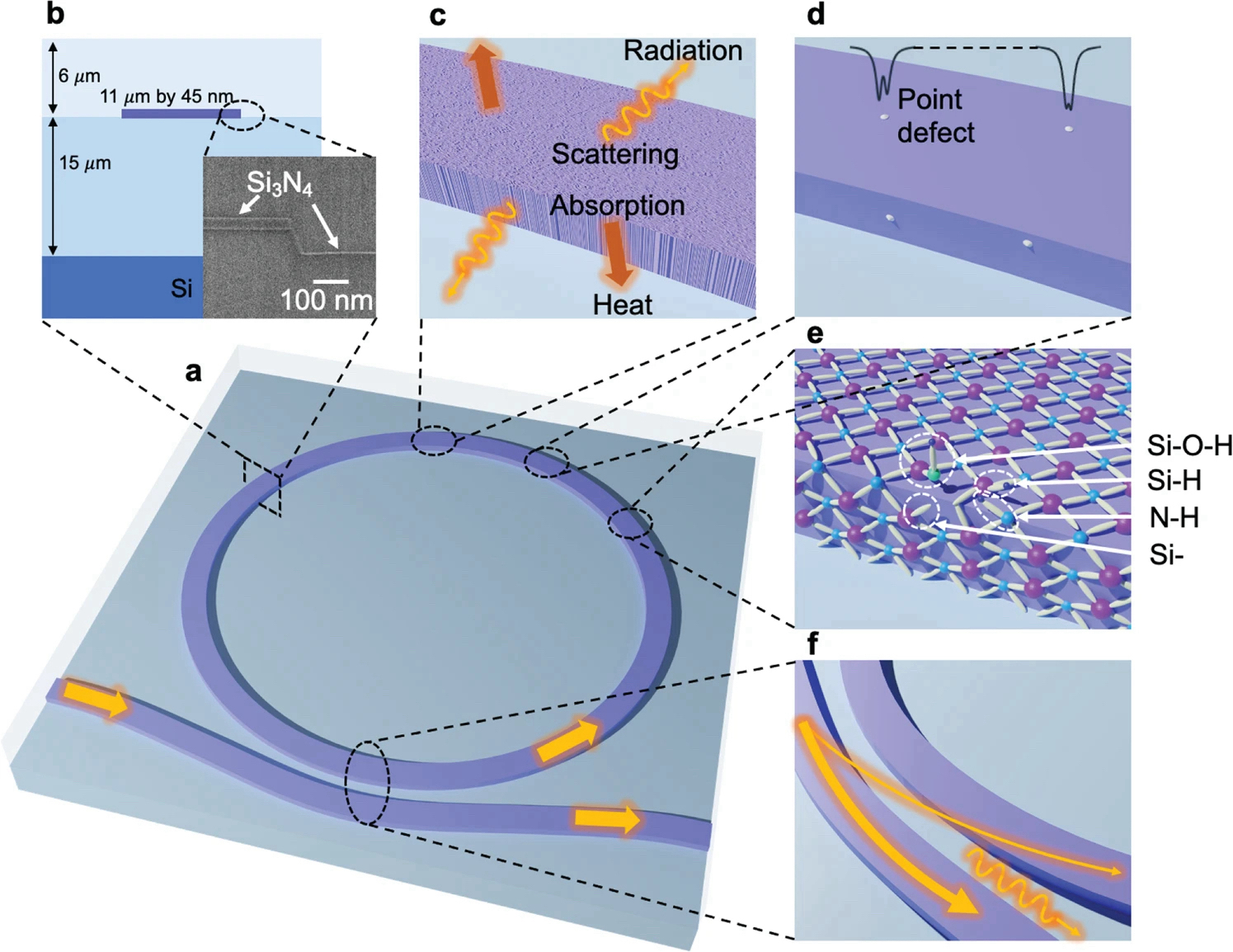}
	\caption{ (a) Illustration of the resonator design. (b) Waveguide design. Inset scanning electron micrograph (SEM) image of the waveguide cross section with the thermal oxide bottom cladding, an etched low-pressure chemical vapor deposition (LPCVD) grown silicon nitride waveguide core, an additional thin (~5 nm) blanket layer of silicon nitride (covering the core top and sidewalls), and a tetraethoxysilane pre-cursor plasma-enhanced chemical vapor deposition (TEOS-PECVD) oxide top cladding. SEM length scale bar is 100 nm. (c) Waveguide surface roughness scatters the guide mode energy into radiation mode and bulk absorption generates heat. (d) Point defects split resonances. (e) Defect bonds such as Si–O–H, Si–H, N–H, and dangling bonds lead to surface absorption (f) The bus-ring coupler scatters energy into radiation modes and adds excess loss. Figures reproduced from Puckett et al.\cite{Puckett2021} under a Creative Commons license \href{http://creativecommons.org/licenses/by/4.0/}{http://creativecommons.org/licenses/by/4.0/}}
	\label{fig1}
\end{figure}

The biggest challenge silicon platforms possess to quantum photonic applications is the photon losses, owing to propagation losses and two-photon absorption (TPA). An approach to reduce these losses is to interface silicon with SiN, which can reduce propagation losses to $<0.1$ $dB m^{-1}$. SiN also eliminates TPA due to its higher bandgap \cite{Puckett2021}. Rosenfeld et al. have recently demonstrated suppression of TPA within silicon by operating in longer wavelengths \cite{Rosenfeld2020}. \\

\section{Lithium Niobate based Platform}\label{sec3}

Lithium niobate (LN) is an emerging platform for quantum photonic integrated circuits (QPIC). LN has often been dubbed as the \textit{silicon of nonlinear optics} or \textit{silicon of photonics}\cite{1_silicon_of_photonics}.  
It was the workhorse of guided wave optics devices like Mach-Zehnder modulators, polarization controllers, and switches even before the advent of silicon photonics \cite{2_workhorse}. LN is a versatile platform because of its broad transparency range from 350 nm to 5200 nm, strong electro-optic, acousto-optic, and thermo-optic effects which make it an excellent choice for active and non-linear devices as well \cite{3_LN_props, 4_LNprops}. Contrary to silicon, LN is an anisotropic material and its anisotropy can be exploited for a multitude of applications. Its birefringence and ferroelectricity facilitate phase matching of the desired nonlinear processes \cite{5_biregringence}. High second-order nonlinearity, electro-optic properties and ability to operate at low temperatures make LN compatible with quantum communication applications \cite{6_quantumCompatible}. The high $ \chi ^{(2)}$ nonlinearity in LN due to lack of centro-symmetry allows for spontaneous parametric down conversion (SPDC) to generate quantum light, where one pump photon is split into a pair of signal and idler photons sharing a common quantum state\cite{7_primaryRefernece}.\\

Traditionally, waveguiding regions in bulk LN are formed by titanium diffusion \cite{8_Tidiffused} or proton exchange \cite{9_protonXchange} that locally increases the refractive index of the core. The induced index change is small, 0.001 to 0.01 and hence these waveguides have weak confinement and large mode size \cite{10_weakConfine}.  With the advent of thin film lithium niobate on insulator (LNOI) similar to SOI, there has been a renewed world-wide interest in considering this material for photonics. The optical mode size in the LNOI waveguides is typically reduced by more than an order of magnitude compared to their bulk counterparts \cite{11_modesize}, which favours compact photonic devices and dense quantum optical circuits.  One disadvantage of LN is that it is a difficult material to etch \cite{12_etchDifficult}, and the roughness of the etched sidewalls as well as re-deposition of the chemical-etching by-products contribute to large amounts of scattering and propagation losses \cite{13_sidewallRoughness}. However, researchers have recently succeeded in etching LN resulting in low loss, high quality waveguides. Fabrication of single-mode LN strip waveguides with optical patterning and chemo-mechanical polishing demonstrated in \cite{14_lowLossWaveguides} have a very low propagation loss of 4.2 dB/m, while the dry etched rib waveguides in \cite{15_lowlosswaveguides} exhibit a propagation loss of 2.7 $ \pm $  0.3 dB/m. LN also offers an added advantage of low loss over a broader spectral range.\\

Even though silicon photonics can shrink the size of photonic devices and achieve highest miniaturization, its merits are diminished by dependence on carrier dynamics for modulation and excess optical loss that results when silicon is doped. This limits its use for high throughput modulation \cite{16_silicon_demerits}. Silicon also lacks second order non-linearity and hence non-linear interactions have to rely on much weaker third order effects. This hinders the performance of optical components like switches and frequency converters \cite{17_siliconDemerits}.\\

The implementation of scalable on-chip quantum circuits depends on various functional blocks which involve the generation, manipulation, detection of photons.\\

LN has the advantage of second-order non-linearity, and hence uses spontaneous parametric down conversion (SPDC) instead of spontaneous four-wave mixing used in silicon and silicon-nitride based photonic circuits \cite{18_fourwaveMixing}. The pump power needed to generate photon pairs with SPDC is lower, but the phase mismatch needs to be taken care of. SPDC-based devices enjoy stronger nonlinear optical effects while not subject to undesirable side $ \chi^{(3)} $ processes. These devices are useful for quantum photonics applications like photon-pair generation, heralded single-photon creation and quantum frequency conversion \cite{19_heralded}.\\

Direct generation and detection of high-purity photon pairs across the visible/near-IR (780 nm) and mid-IR (3950 nm) spectra using SPDC in a periodically poled lithium niobate (PPLN) waveguide is reported in \cite{20_directGeneration}. Such highly non-degenerate photons can be used to synthesize disparate quantum nodes and link quantum processing over incompatible wavelengths \cite{7_primaryRefernece}. Photon-pair generation at high rates of 8.5 and 36.3 MHz using only 3.4 and 13.4 $\mu W$ pump power, respectively, using a periodically poled lithium niobate microresonator on chip has been demonstrated in \cite{20_directGeneration}. This work can be used to implement high-dimensional entangled quantum states and optical quantum logic as high-quality single and correlated photons are created over multiple wavelength channels simultaneously.
For the generation of quantum-correlated photon pairs, it is necessary to filter out the pump signal. LNOI waveguides for filtering application using lateral leakage between the waveguide modes and slab modes of orthogonal polarization has been demonstrated in \cite{21_LN_filters}. This enables the periodically-poled LN waveguides generating photon pairs and subsequent filtering section where the pump field leaks into the slab.\\

The exceptionally high second-order non-linearity of LN has been harnessed via quasi-phase matching to demonstrate a periodically poled thin film lithium niobate microring resonator that reaches 5,000,000$ \% $/W second-harmonic conversion efficiency\cite{22_secondHarmonic}. A single-photon anharmonicity FOM of 0.7 $ \times $ 10$ ^{-2} $ is obtained in this work. The authors also theoretically propose device configuration that allows for single-photon filtration via unconventional photon blockade effect. This could pave the way for emitter-free, room-temperature quantum photonic applications, such as quantum light sources, photon–photon quantum gate, and quantum metrology.\\

Quantum networks consist of quantum channels along which photonic qubits travel and atomic ensembles which are used to store and process the qubits \cite{23_qubits}.  This arises a need for quantum information interfaces to transfer qubits from one type of carrier to another. For photonic carriers, the interface needs to provide wavelength adaptation to atomic levels while preserving quantum coherence of the initial state. Such interfaces have already been reported in soft-proton exchange: PPLN waveguides with the single-photon up-conversion transfer based on mixing the initial 1312 nm single photon with a highly coherent pump laser at 1560 nm \cite{23_qubits}.   Frequency conversion of single photons across spectral bands is desirable for hybrid quantum systems connecting disparate nodes, like cold atoms and superconducting Josephson junctions, where the photons from each source must be transduced into indistinguishable quantum states \cite{24_quantumStatesNoise}. It is also necessary to facilitate the up-conversion detection of near-infrared single photons for low cost and high detection efficiency. A noise analysis of photon conversion in thin-film LN waveguides have been carried out in \cite{24_quantumStatesNoise} and reports the noise level on the order of $ 10^{-4} $ photons per time-frequency mode for high conversion. The generation of spectrally unentangled biphoton states using LN photonic crystal slab waveguides have been demonstrated, which is highly desired for heralding of single photons in \cite{25_biphotonStates}.\\

Quasi-phase-matched frequency conversion in a chip-integrated lithium niobate microring resonator, with a normalized efficiency of $  10^{-6} $ per single photon is reported in \cite{26_phasematching}. Such quasi-phase matching and electric field poling in LN waveguides enables frequency conversion between arbitrary wavelength bands. The authors also mention that with further increase in Q-factor, non-linearity can reach single-photon level, which allows deterministic entanglement generation and control-NOT gate for single photons. This paves the way for quantum light generation leading to scalable, nonlinear-optical quantum computing.\\

Electro-optic (EO) tuning is an energy efficient method for achieving complex filtering and routing functions in photonic integrated circuits for WDM networks. Such EO tuning is demonstrated for micro-ring resonators with a tuning range of 1.38 nm \cite{27_EO1}. EO effect can also be used to build reconfigurable quantum gates. Modulators on LN with an ultra-wide 3 dB bandwidth exceeding 400 GHz has also been reported \cite{28_EO2}. Compact LN modulators with an ultra-low half-wave voltage ($ V _{\pi} $) of 1.08 V have already been achieved on an 8-mm-long modulator containing three-segments of 7-mm long modulation sections, together with a 3-dB electro-optic bandwidth around 43 GHz \cite{29_EOM}. Such low Vpi combined with a compact footprint has the potential to be directly driven by CMOS electronics. An electro-optic amplitude modulator with a sub-1-volt half-wave voltage of only 0.875 V and an interaction region length of 2.4 cm is also reported in \cite{30_subvolt_modulator}. Phase modulators at the heart of amplitude modulators play a key role in the implementation of quantum cryptographic schemes like differential phase shift quantum key distribution \cite{31_DPSK}.\\

LN platform can be further enriched by incorporating rare-earth ions, so that it finds new applications in quantum information processing. Rare-earth ions are potential candidates for quantum information processing, quantum memory protocols for quantum networks, light-matter interactions, and quantum-state teleportation \cite{32_Erdoped}.  Optical properties at 1.5 $ \mu $m wavelengths of rare-earth ions ($ Er^{3+} $) implanted in thin-film lithium niobate waveguides and micro-ring resonators are investigated in\cite{32_Erdoped}. A platform for rare earth ions on thin film LN has also been demonstrated \cite{33_rareEarth} which can open new avenues for scalable compact lasers, and on-chip quantum memory. An ytterbium-doped lithium niobate microring laser operating in the 1060-nm band under the pump of a 980-nm-band laser is realized in \cite{34_ytterbium}. This monolithic laser has a low threshold of 59.32 $ \mu $W and relatively high output power of 6.44 $ \mu $W, a state-of-the-art value for rare-earth ions-doped lithium niobate thin-film lasers. A C-band wavelength-tunable microlaser with an Er3C-doped high quality ($ \sim1.8 \times 10^{6} $) lithium niobate microdisk resonator is also demonstrated in \cite{35_microdisk}. \\

LN is a negative uniaxial material with ordinary index ($ n_{or} $) along two of the principal axes and, extraordinary index ($ n_{ex} $) along the optic axis of the crystal. This anisotropy offers another degree of freedom in terms of polarization, which can be used for spatial routing of photons.  The information capacity of qubits can be increased with the use of multiple degrees of freedom of a quantum particle simultaneously. These include polarization, frequency, time, orbital angular momentum for photons \cite{36_multipleDegrees}. The polarization beam splitter (PBS) is one of the most important polarization-handling devices for splitting and combining the transverse-electric (TE) and transverse-magnetic (TM) polarizations \cite{37_PBS}. Interplay of waveguide birefringence and intrinsic birefringence of LN can be used to design dual-mode waveguides and directional couplers that are polarization-independent with no modal birefringence \cite{38_ZBR}. \\

Interaction between photons and energy transfer is essential to implement scalable quantum optical systems using quantum gates. Such interaction is achieved using non-linear processes like sum-frequency generation. An ultraefficient sum-frequency generation on a chip using a fully optimized periodically poled lithium niobate microring, with an external quantum efficiency of (65 $ \pm $ 3)$\% $ with only (104 $ \pm $ 4) $ \mu $W pump power has already been demonstrated \cite{39_sumfreq}. The authors also mention the feasibility of the demonstrated photon conversion and interaction to be integrated with other passive and active elements on the same chip, such as PPLN waveguides, electro-optical modulators, frequency-comb sources, and microring filters to create functional quantum devices.\\

Large-scale quantum networks also necessitate optical quantum memories for the storage of photonic quantum states to synchronize different functionalities on a chip \cite{7_primaryRefernece}, apart from generation and interaction of photons. Long waveguides with low loss can serve as simple storage elements with storage time decided by the propagation delay. Waveguides on LN have the additional benefit of tunability through electro-optic effect. Such electro-optically switchable optical true delay lines on lithium niobate on insulator using photolithography assisted chemo-mechanical etching are illustrated in \cite{40_EOdelay}. The reported device generates different amounts of time delay as the configuration consists of several waveguides of different lengths which are consecutively connected by electro-optical switches. A time delay of 2.2 ns was measured with a delay line of 32 cm and propagation loss remained low ( $ \sim $0.03 dB/cm) for waveguide lengths well above 100 cm. Another implementation demonstrates coherent control if frequency and phase of light in a ‘photonic molecule’ with two distinct energy levels using coupled lithium niobate microring resonators \cite{41_photonicMolecule}. By reconfiguring the photonic molecule into a bright–dark mode pair, on-demand optical storage and retrieval is made possible. This dynamic control of light in a programmable and scalable electro-optic system opens the door for quantum photonic gates in the frequency domain. Doping of LN with rare-earth ions can also open the door for quantum storage devices on this versatile platform.\\

In order to analyze quantum states and to measure the results of photonic-quantum operations, superconducting nanowire single-photon detectors (SNSPDs) are proved to be powerful components. SNSPDs are experimentally demonstrated on thin-film LN with an on-chip detection efficiency (OCDE) of 46$ \% $, a dark count rate of 13 Hz, a timing jitter of 32 ps. However, the efficiency needs to be improved further to compete with other platforms such as Si (91$ \% $) and SiN (70$ \% $) \cite{42_LNSPDC}. The monolithic integration of electro-optically reconfigurable LN circuits with superconducting nanowire single-photon detectors (SNSPDs) paves the way for realizing scalable photonic devices for active manipulation and detection of quantum states of light. Joint cryogenic operation of SNSPDs and an electro-optic modulator is demonstrated in \cite{43_SNSPD}. The device consists of a balanced and tunable Mach–Zehnder interferometer (MZI) made of an electro-optic phase shifter and two waveguide-integrated SNSPDs at the two outputs. Figure 1 shows a microscope image of the integrated device employed for demonstrating the joint cryogenic operation of SNSPDs and an EOM, along with a schematic of the experimental setup \cite{43_SNSPD}.
\begin{figure}[ht]
	\centering
	\includegraphics[width=0.9\linewidth]{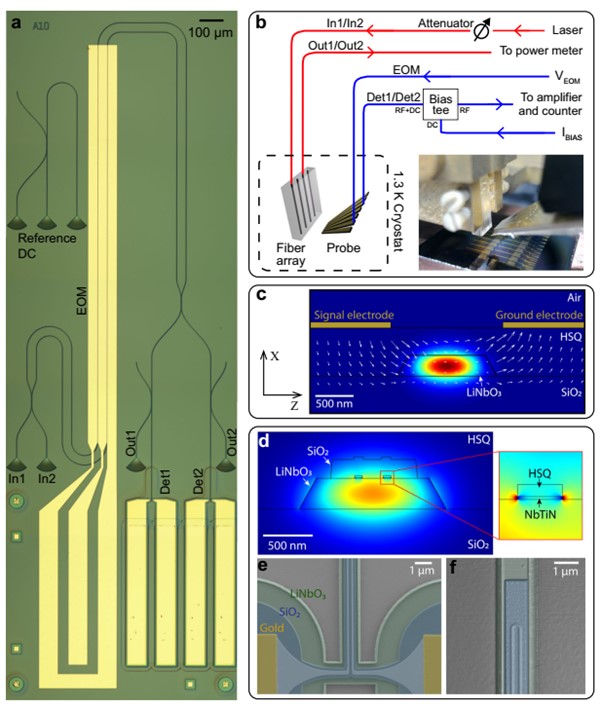}
	\caption{ (a) Optical microscope image of the integrated device (b) Schematic of the experimental setup with optical (red) and electrical (blue) access to the LN-chip. Inset in the figure shows a photograph of the photonic chip, the fiber array, and the contact probe employed in our experiment. (c) Electrode configuration used for the realization of the electro-optic phase shifter (d) Schematic of the nanowire configuration used for the realization of waveguide-integrated SNSPDs, and simulation of the nanowire absorption. (e) False-color scanning electron micrograph of a waveguide-integrated SNSPD taken in the proximity of the gold electrodes pads (f) Surface topography of a waveguide-integrated SNSPD taken in the proximity of the U-turn by atomic force microscopy. Figures reproduced from Lomonte et al.\cite{43_SNSPD} under a Creative Commons license \href{http://creativecommons.org/licenses/by/4.0/}{http://creativecommons.org/licenses/by/4.0/}}
	\label{fig2}
\end{figure}
\\

Coupling of light from integrated-optic waveguide to optical fibers is necessary to transfer photons into networks, or between separate on-chip subsystems. Large-scale quantum computing architectures and systems may ultimately require quantum interconnects to enable scaling beyond the limits of a single wafer, and toward multi-chip systems. Coupling light from a single mode fiber to LNOI devices is difficult due to the large mode mismatch between the single-mode fiber and the submicron-LNOI waveguide \cite{44_Coupling}. Grating couplers are a viable solution for coupling light from a waveguide to an optical fiber. Monolithically defined taperless grating couplers in Z-cut lithium niobate on insulator for efficient vertical coupling between an optical fiber and a single mode waveguide is demonstrated in \cite{45_gratingCoupler}. These grating couplers in C/L band exhibit $\sim$ 44.6$\%$/coupler and $\sim$ 19.4$\% $/coupler coupling efficiency for TE and TM polarized light respectively. Self-imaging apodized grating couplers without metal back-reflector, featuring a coupling efficiency 47$ \% $ for TE0 mode at $ \lambda $=1550 nm and 44.9$ \% $ for TM0 at $ \lambda $=775 nm is reported in \cite{46_selfImaging}. A peak coupling efficiency of 78$\% $ for z-cut LNOI around 1550 nm with a 3-dB-bandwidth of 98 nm was achieved for TE polarization with the use of gold layer as a bottom reflector \cite{47_bottomReflector}.\\

To summarize, lithium niobate meets all the requirements of a quantum photonics platform as listed in  \cite{Moody2022}: (i) has ultra-low loss to preserve fragile quantum states; (ii) enables precise control of the temporal and spectral profiles of photons; (iii) allows fast and low-loss optical switches to route quantum information; (iv)  able to operate in visible and telecom wavelengths (v) features strong nonlinearities for efficient frequency up- and down conversion, quantum transduction, and entangled photon pair generation; (vi) allows integration of photodetectors and operating electronics. Commercial availability of LNOI wafers resulting from advanced fabrication technology along with its versatile material properties make LN an indispensable material for the implementation of quantum integrated optic circuits. Also, monolithic integration of LN devices on silicon substrate makes it compatible with silicon photonics. LN can also be heterogeneously integrated with silicon, so that the benefits of both the cornerstone materials can be leveraged \cite{49_hybridSilicon}. This combination can exploit the features of both the platforms, like the high index contrast and easy etching of SOI, and large electro-optic coefficient of LNOI. Hybrid silicon and lithium niobate Mach-Zehnder modulator operating at C-band and O-band simultaneously with a bandwidth exceeding 70 GHz is already demonstrated \cite{50_hybridSiliconMod}. The authors of \cite{7_primaryRefernece} also predict LNOI platform could facilitate filtering of pump with an extinction ratio exceeding 100 dB.\\

\section{Diamond based Platform}\label{sec4}

The diamond's outstanding mechanical, thermal and optical properties, once recognised, made it a suitable alternative for integrated photonics-based research, unlike its traditional use as gemstones \cite{doi:https://doi.org/10.1002/9783527648603.ch1,Mi2020}. Silicon dominates the commercial market regarding on-chip photonics chips for telecom, spectroscopy, and quantum applications \cite{JEREMY}.
The demand for compatible single photon sources for quantum registers and quantum memories can be quenched by the inherent properties of a diamond based platform for quantum information processing applications \cite{Dutt2007, Doherty2013, Wehner2018, Atature2018}. Quantum coherent frequency conversion and quantum memories can come together on the same platform utilising the diamond's excellent opto-mechanical properties. Even though research has been going on for a single component, no material has been discovered which has the potential to become an \textit{all in a single integrated device} \cite{Awschalom2018, Sipahigil2016}. Single Crystal Diamond (SCD) applications have been demonstrated in communication, quantum spectroscopy, and deep-UV to mid-infrared Raman \cite{Granados2011, Latawiec2018, Williams2018}. \\

Growth of commercial SCD is mostly done either by high-pressure high temperature (HPHT) method or by homoepitaxial chemical vapour deposition (CVD) \cite{Tatsumi2018, Silva2009} Even though the HPHT method shows inhomogeneously distributed impurity, low dislocation density makes it unsuitable for wafer production \cite{Terentyev2015}. Homoepitaxial CVD has shown better quality wafer growth by properly tuning growth chemistry and impurity control. Still, the height of the substrate grown is limited \cite{Balmer2009}. However, heteroepitaxial CVD has shown higher yield properties and almost matchable impurity and other properties than homoepitaxial CVD \cite{2018219, 20181}. To make it optoelectronic compatible, thin films of the SCD have been demonstrated by thinning down the bulk. The challenges are achieving the required smoothness and uniformity, and the stack of 5 $\mu$m is available in the market. The patterning of SCD or thin-film diamond is structured with e-beam lithography. A wide range of etching techniques has been explored, like wet etching, metal catalytic etching, dry etching, and plasma etching, exploiting its chemistry with subsequent etchants \cite{Atikian2017, Naamoun2012, Tallaire2016, Mouradian2017}.\\

Passive devices like waveguides, couplers, resonators, splitters, combiners, and modulators fabricated on diamond surfaces are reported. Active devices like single photon sources, supercontinuum lasers, Raman lasers, colour centre lasers and detectors like superconducting nanowire single photon detectors (SNSPDs) are also being reported \cite{PhysRevApplied.8.024026, PhysRevX.9.031022, Kato2013}. Total monolithic integration on a single substrate needs improvement and precision in each fabrication step. 

\section{III-V Group Materials based Platform}\label{sec5}

III-V materials such as InP or GaAs and devices based on their compounds are a rapidly growing platform because of their direct bandgap and light-emitting properties, which are explored to make active devices such as lasers, LEDs and modulators. The ability to integrate them with passive devices has felicitated the rapid progress with quantum emitters based on InAs quantum dots (QDs). Single-chip integration is possible in two ways, either by monolithic or heterogeneous integration, and both can be done in batch processing. \\

Epitaxial growth of III-V on silicon is one of the earliest fabrication methods which rapidly grows with the field of silicon photonics. QDs have proven to be more resilient to crystalline defects since the carriers are localized to individual dots provided the density of QD is much greater than the density of crystal defect. This makes QD lasers more attractive as light sources though their lifetime and reliability are dependent on the threading dislocation density (TDD). Lifetime over $10 \times 10^6$  hours have been achieved at room temperature with a TDD of $7 \times 10^{6} cm^{-2}$ by Jung et al. \cite{Jung2018}. The same group had earlier reported microring lasers with quantum dot active regions on silicon with 1 mA threshold current and lasing upto $100^{o}C$ while reducing the threshold current density to $200 Acm^{-2}$. Lasers on GaP/Si templates have been demonstrated with single-facet output powers more than 175mW and efficiencies upto 38.4\% \cite{Jung2017, Jung2017b}. A wide range of methods including quantum dot intermixing, etching and regrowth and offset quantum dot design have been propsoed for active passive integration as modulators and photodetectors based on III-V materials are also grown in silicon. Further research is needed to decrease TDD so that QD lasers have better reliability and longer life at higher temperatures. Another area which requires further improvement is coupling between epitaxial QD lasers with silicon waveguides and components. The thickness of buffer layers between the QDs and silicon make optical coupling a challenging task. Various methods have been proposed including growing III-V in pre-etched trenches with active region vertically aligned with the silicon waveguide or reducing the buffer layer thickness by aspect ratio trapping in nanoscale trenches. Researchers are also pursuing a platform completely made of III-V materials using the silicon wafer only as substrate  \cite{Bowers2019}.\\

Heterogeneous integration is the process of bonding two different types of materials followed by device fabrication in a single process flow. This approach is used to fabricate optoelectronic and nonlinear photonic devices. Especially the devices with enhanced optical nonlinearities provided by III–V semiconductor materials on silicon substrates and on-chip waveguides \cite{Bowers2019, Moody2020}. Bakir et al. used this method for the first time and produced optically pumped lasers based on III–V epitaxial QDs on a silicon substrate \cite{Bakir2006}. Tanabe et al. later demonstrated electrically injected QD lasers on silicon using the same method \cite{Tanabe2010}. Heterogeneously integrated single photon sources in SiN-based photonic circuits at chip-scale were recently demonstrated by  Davanco et al. \cite{Davanco2017}. Schnauber et al. deterministically fabricated single photon sources with QDs precisely in GaAs nanowaveguides. They demonstrated post-selected indistinguishable single-photon emission into SiN waveguides. The coherence lengths were of the order of those obtained from InAs self-assembled QDs in GaAs-only devices \cite{Schnauber2019}. Introducing single QDs in quantum silicon photonic circuits can considerably scale integrated quantum photonic information systems relying on probabilistic gates. Single QDs can act as on-chip sources of indistinguishable photons that can efficiently couple with silicon photonic circuits through GaAs nanophotonic structures with high success rates. They also manifest single-photon nonlinearities that can enable on-chip deterministic quantum gates \cite{Kim2013}. On-chip QD single photon sources are now produced in silicon photonic devices by the same hybrid integration techniques in which  III–V and silicon-based devices are produced \cite{Elshaari2020}.\\

\begin{figure}[ht]
	\centering
	\includegraphics[width=0.9\linewidth]{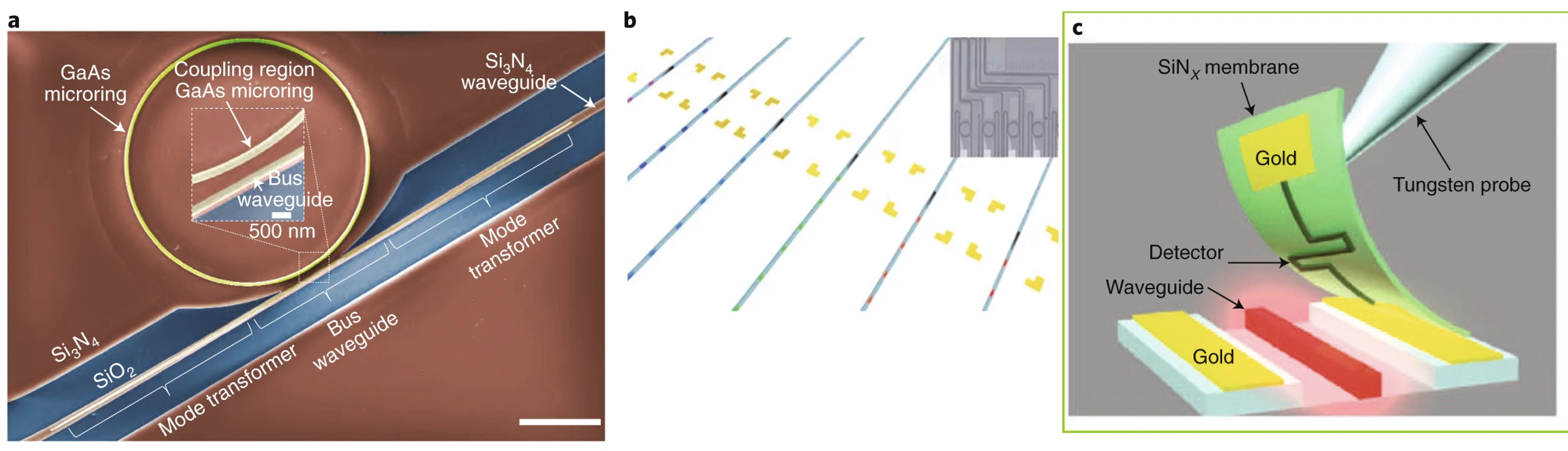}
	\caption{ (a) Photon Source - Wafer bonding of a gallium arsenide ring resonator with QDs to silicon nitride waveguides (scale bar, 10 $\mu$m). (b) Photon Source - Encapsulation of multiple nanowire QD single-photon sources in silicon nitride waveguides. (c) Detector - Hybrid integration of SNSPDs fabricated on silicon nitride membranes on aluminium nitride photonic waveguides using a pick-and-place technique. Figures reproduced from Elshaari et al.\cite{Elshaari2020} under a Creative Commons license \href{http://creativecommons.org/licenses/by/4.0/}{http://creativecommons.org/licenses/by/4.0/}}
	\label{fig3}
\end{figure}

An interface with higher efficiency is required between a single QD and a spatially confined optical mode single that should be efficiently accessible.
To maximise coupling, the QD has to be precisely positioned in with sub-wavelength-scale at specific locations near the spatial optical mode, which is challenging in terms of scalable integration. This has been achieved to an extent by site-selective QD growth techniques. Stranski–Krastanow (S–K) growth mode can be used to generate self-assembled QDs with greater coherence, which is critical for quantum photonic application. Nevertheless, this is still a challenging area and a field of active research \cite{Thomas2008, Moody2022}. \\

\section{Rare Earth Metals based Platform}\label{sec6}

Rare earth elements or rare earth metals are 17 chemical elements in the periodic table with similar chemical properties. That includes the 15 lanthanides, scandium and yttrium (which tend to occur in the same ore deposits as the lanthanides). Despite their classification, most of these elements are not \textit{rare}. These elements are essential in consumer electronics, computers and networks, communications, clean energy, advanced transportation, healthcare, environmental mitigation, and defence technologies. Since many rare-earth doped materials simultaneously possess long-lived spin coherence and stable optical transitions, they have gathered significant interest and emerged as a rapidly advancing field in recent years \cite{Zhong2019}. They offer a vast range of possibilities for quantum device engineering with efficiency and scalability, paving the way for a new paradigm called Rare Earth Quantum Computing (REQC). In 2018, the European Union announced a flagship project called SQUARE (Scalable Rare Earth Ion Quantum Computing Nodes) with a budget of 3 million euros, with a mission to establish rare earth ions as a fundamental building block of a quantum computer and to overcome the main roadblocks on the way towards scalable quantum hardware. \\

Rare earth metals are commonly used as dopants in crystals like yttrium aluminium garnet (Y\textsubscript{3}Al\textsubscript{5}O\textsubscript{12} or YAG), Y\textsubscript{2}SiO\textsubscript{5} (YSO), Y\textsubscript{2}O\textsubscript{3} or YVO\textsubscript{4} (YVO). The rare earth ions' 5s and 5p outer orbitals shield the electrons in the partially occupied 4f orbitals from interacting with the crystalline environment. The electrons in 4f orbitals are highly localised. In free space, 4f–4f transitions are parity-forbidden, but in the presence of a crystal field, they are weakly possible, which gives rise to sharp optical transitions with high quantum efficiency. At cryogenic temperatures sharp optical transitions with Q-factor over $10^{11}$ are observed with long optical $T_1$ lifetimes of order hundreds of $\mu$s to ms. \\

Optical quantum memory is one area where rare earth doped crystals have proven more advantageous than other platforms. In a rare earth doped material platform, qubits encoded on single photons can be stored better than nitrogen-vacancies in diamonds and semiconductor quantum dots in decoherence time and read-out. Ensemble-based quantum memories have been demonstrated in rare earth metal based platforms, including storage of single photons, photonic quantum entanglement, and quantum spins \cite{Clausen2011, Zhong2017, Jobez2015}. They also have the advantage of being natural interfaces between optical photons and spins in RF and microwave frequencies. This property can make quantum transducers interconnect with distant superconducting quantum circuits. Rare earth material based hybrid quantum systems also holds promise for microwave quantum storage and microwave-to-optical conversion. Latest developments in this direction include the demonstration of high cooperativity coupling of rare-earth spins to a superconducting resonator and experimental implementation of microwave to optical conversion via magneto-optic coupling with maximum efficiency \cite{Zhong2019, Wolfowicz2015, Williamson2014, Xavier2019}. \\

In REQC, nuclear spins of the ions are used to code qubit states, and optically excited states ate used to perform gate operations. Frequency addressing of qubits which are nm apart can be achieved through this. Single qubit gates with fidelity errors of the order of $10^{-4}$ have been designed, but further research is needed to attain better fidelity. Dipole blockade (the electrical or the magnetic dipole-dipole interactions) is used for entanglement generation in REQC. For two qubit gates, fidelity errors are of the order $10^{-3}$ since they consist of several single-ion operations. Novel gate protocols are required to improve fidelity.\\

Further developments in the micro-cavity technology may allow new methods for quantum entanglement utilising the cavity mode, increasing the gate fidelity and the qubit connectivity. For qubit readout in REQC, the best method is when several qubit ions are connected to a separate readout ion Purcell enhanced via cavity \cite{Wang2019}. This leaves excited states of qubit ions unaffected and maximises the readout fidelities. Theoretically, high emission rates can be achieved by different types of cavities, which can be chosen according to their characteristics like the ability to scan to multiple spatial spots or the ability to tune in the ns range and ability for on-chip integration. We can use the dipole-blockade mechanism to sequentially find which ions can control the readout ion and which ions can control them. Thus, a set of connected ions can be used to realise a Noisy Intermediate Scale Quantum (NISQ) processor node. One NISQ qubit node with a readout ion in a cavity can be connected to the readout ion in another cavity by detecting interfering photons. High entangling fidelity can be maintained by doing this in a heralded way, similar to scaling in other schemes for quantum computing. RE ions trapped in nano-materials are excellent candidates for integrated photonic platforms. They can function at the telecom wavelengths and are scalable due to the ability for miniaturisation and good photon connections \cite{Kinos2021}.\\

The most popular material among rare earth elements is the Y\textsubscript{2}SiO\textsubscript{5} crystal as a host doped with Eu3+ ions as qubits at a few percent levels. There can be races of other RE ions at $<$1 ppm, and Nd or Er could be used as readout ions. It could be produced as a thin film for future scalable production, and ions can be implanted deterministically. The limitations include low $T_2$, and further research is needed for good optical $T_2$ and readout. \\

\section{Future Trends}\label{sec7}

We have surveyed major electronic materials based integrated photonic platforms that enable quantum information processing. However, this is not an exhaustive list. Developments of new materials and experimental paradigms are an ongoing area of research.    \\

There are new materials like tantalum pentoxide or tantala (Ta\textsubscript{2}O\textsubscript{5}), which have found generated interest in quantum technology researchers in recent years due to many interesting optical properties. They exhibit propagation loss of only 3 dBm\textsuperscript{-1} on oxidised silicon wafers in CMOS compatible fabrication processes due to their high refractive index contrast and low optical absorption. They have wideband transparency for wavelengths from 300 nm to 8 $\mu$m and a third-order nonlinear Kerr coefficient, three times that of SiN. Their large bandgap (3.8–5.3 eV) outweighs the lower nonlinear refractive index compared to silicon photonic devices. This has already led to the demonstration of Ta\textsubscript{2}O\textsubscript{5} micro ring resonators at telecom wavelengths with several million optical quality factors. Ta\textsubscript{2}O\textsubscript{5} is a good choice for realising passive, active, and nonlinear integrated photonic devices. The ability of Ta\textsubscript{2}O\textsubscript{5} to host rare-earth ions like Erbium also allows it to be used as a gain medium that can be used for on-chip optical signal amplification and lasing capabilities \cite{Moody2022, Subramani2010}. However, much further research is needed on the material before it can be projected as a viable solution as an integrated photonics platform.\\

The technological advances in integrated photonic platforms for quantum technologies are not restricted to materials alone. New paradigms like machine learning and quantum technology beyond qubits are the latest trends in the field. Machine learning algorithms have generated interest in photonics primarily due to their ability to reduce the computational cost since they can automatically learn patterns within a dataset without a physical model. These algorithms significantly increase the speed of forward design and enable high-speed inverse design. The type of machine learning algorithms like discriminative artificial neural networks specifies nonlinear mappings between input and output variables. They use matrix multiplication, neuron-based processing, and nonlinear activation functions to compute forward problems faster than traditional electromagnetic solvers \cite{RevModPhys.91.045002, Hammond2019}. These techniques have found applications in the design of integrated photonic platforms, ultrafast photonic systems and quantum technological applications \cite{Jiang2021, Genty2021}. Techniques are developed to solve Schrodinger’s equation rather than Maxwell’s equations and encode an exponential number of terms describing a quantum state in a neural network with a polynomial number of qubits enabling efficient quantum state tomography \cite{Dunjko2018, Tiunov2020}. The field of artificial intelligence combines with quantum photonics to open a plethora of possibilities. It can be used to suggest new quantum experiments and also to develop photonic hardware to train and implement machine learning models \cite{Krenn2020}. However, there is a need to improve optimization techniques, and strategies before quantum photonic implementations of neural networks can improve enormous multi-dimensional datasets generated by quantum computers \cite{Moody2022}. \\

Qudits are \textit{d}-level systems of the more ubiquitous two-level qubits. They utilise higher dimensionality, which is an inherent property of quantum systems. They outperform qubits by providing access to a larger state space for storing and processing quantum information, enhancing robustness against eavesdropping and simplified gates. With the demonstration of qudit entanglement systems overcoming noise in entanglement distribution, they are considered the future of quantum technologies in communication, sensing, and computation \cite{PhysRevX.9.041042}. High room-temperature coherence times, ease of transmission in free space and optical fibre, and efficient detectors make photonic platforms excellent for implementing quantum technologies based on qudits. Recent developments in on-chip qudits have made this technology a burgeoning field. Wang et al. demonstrated a 15 × 15 dimensional entanglement using a silicon photonic device that integrated photon-pair sources, multimode interferometers, phase-shifters, beamsplitters, waveguide crossers, and grating couplers \cite{Wang2018}. Quantum teleportation between two silicon chips have been demonstrated by Llewellyn et al. \cite{ada6d0b20e644f3199ca99b1c2097faf}. Kues et al. and Imany et al. showed two-qudit gates acting on photons generated via SFWM in a microring resonator \cite{Kues2017, Imany2019}. Using the transverse spatial modes, an on-chip source of polarisation and OAM modes has been demonstrated on a silicon platform by Mohanty et al. \cite{Mohanty2017}. They have shown quantum interference using photons generated from SPDC and SiN waveguides that act as mode multiplexers and beamsplitters. Photonic qudit information processing on the chip is an active area of research, with path, time-bin, and frequency encodings as the most developed in experiments. The challenge for qudit technology lies in scalability. It increases significantly when quantum error correction is employed. Another challenge is the implementation of universal quantum gates. For path encoding techniques, thermal phase shifting is a big challenge. For frequency modes, conversion is often achieved using nonlinearities, and the challenges here lie in achieving high selectivity and efficiency. Recent results using LN showing 99\% efficiency with programmability indicate promise in this direction since its compatible with optical fibre networks in terms of frequency \cite{Hu2021}. However, qudit-based technology needs further research and development to precisely control and integrate sources, gates, and detectors into one device. Qudit gates are not much developed in these, although recent advances in programmability for frequency and path hold promise for the quantum regime \cite{Moody2022}.\\

In the last few decades, integrated photonic platforms have emerged as a potential technology to make quantum technology the next technological breakthrough. Recent advances in quantum photonics enabled the experiments usually conducted on large quantum optics tables to be scaled down to prototype chips. The coming decades will see extensive scale integration of multi-functional and re-configurable devices that will push the boundaries of quantum information science and engineering. This may require the development of a sustainable quantum photonic ecosystem that can bring together interdisciplinary experts, infrastructure and testbeds, and governmental, academic, and private partnerships. Such an infrastructure should broaden access to quantum integrated technologies and cultivate future generations of a quantum workforce through training and professional development, including modernised teaching labs and hands-on approaches to quantum mechanics and photonics education. \\


\bibliography{sn-bibliography}


\end{document}